\documentclass[prl,aps,twocolumn,showpacs]{revtex4}

\usepackage{epsfig}
\usepackage{amsfonts}
\usepackage{amssymb}
\usepackage{mathrsfs}
\usepackage{theorem}
\usepackage{amsmath}
\usepackage{times}
\usepackage{color}

\usepackage{ifpdf}
\ifpdf
\usepackage{epstopdf}   
\fi

\newtheorem{theorem}{Theorem}
\newtheorem{defin}{Definition}
\newtheorem{lem}{Lemma}

\newcommand{\earl}[1]{\textcolor{black}{#1}}

\newcommand{\ket}[1]{\ensuremath{\vert#1\rangle}}

\newcommand{\kb}[2]{\ensuremath{\vert #1 \rangle \langle #2 \vert}}

\newcommand{\fsa}{f^S_{\vec{a}}}
\newcommand{\fsap}{f^S_{\vec{a}_{P}}}

\renewcommand{\vec}[1]{\ensuremath{\mathbf{#1}}}

\newcommand{\tr}{\ensuremath{\mathrm{tr}}}

\def\unity{\mbox{\small 1} \!\! \mbox{1}}

\def\unity{\mbox{\small 1} \!\! \mbox{1}}

\begin{document}
\title{Bound States for Magic State Distillation in Fault-Tolerant Quantum Computation}

\author{Earl T. Campbell}
\affiliation{Department of Physics and Astronomy, University College London, Gower Street, London, WC1E 6BT, UK.}
\author{Dan E. Browne}
\affiliation{Department of Physics and Astronomy, University College London, Gower Street, London, WC1E 6BT, UK.}

\begin{abstract}
Magic state distillation is an important primitive in fault-tolerant quantum computation. The magic states are pure non-stabilizer states which can be distilled from certain mixed non-stabilizer states via Clifford group operations alone. Because of the Gottesman-Knill theorem, mixtures of Pauli eigenstates are not expected to be magic state distillable, but it has been an open question whether all mixed states outside this set may be distilled.  In this Letter we show that, when resources are finitely limited, non-distillable states exist outside the stabilizer octahedron. In analogy with the bound entangled states, which arise in entanglement theory, we call such states bound states for magic state distillation.
\pacs{03.67.Pp}
\end{abstract}

\maketitle     

The significant noise and decoherence in quantum systems means that harnessing these systems for computational tasks must be performed fault tolerantly~\cite{STEANE95,G01a}. In a wide variety of setups only a limited set of gates, known as the \textit{Clifford group}, are implemented in a manifestly fault tolerant manner.  Examples include some anyonic topological quantum computers~\cite{Moore90,Lloyd02,Doucot02}, post-selected quantum computers~\cite{Knill05,Rei02a} and measurement based topological quantum computers~\cite{RHG01a}.  This motivates the problem of when such devices, with practically error free Clifford gates, may be promoted to a full quantum computer.  The celebrated Gottesman-Knill theorem shows that a Clifford circuit acting on stabilizer states --- simultaneous eigenstates of several Pauli operators --- can be efficiently simulated by a classical computer~\cite{G02a}.  However, given a resource of pure non-stabilizer states, we can implement gates outside the Clifford group.  For example, a qubit in an eigenstate of the Hadamard enables one to implement a $\pi/8$ phase gate that when supplementing the Clifford group gives a dense covering of all unitary operations~\cite{BraKit05}, and so enables universal quantum computation. 

Preparation of non-stabilizer states would usually require  a non-Clifford operation, so in this context, one would require that even noisy copies of these states enable high fidelity quantum computation.  Bravyi and Kitaev~\cite{BraKit05} showed that this can be achieved.  Coining the term \textit{magic state distillation}, they showed that most mixed non-stabilizer states can be \textit{distilled}  via Clifford group circuits to fewer copies of a lower entropy state, reaching in {the limit of infinite iterations a pure non-stabilizer \textit{magic state}.  However, the protocols they presented do not succeed for all mixed non-stabilizer states. Bravyi and Kitaev were not satisfied by the ambiguous status of these states and concluded that ``\textit{The most exciting open problem is to understand the computational power of the model in [this] region of parameters.}''.  Either all non-stabilizer states are efficiently distillable by an undiscovered protocol, or there exist non-stabilizer states that are impossible to distill.  Such undistillable states we call \textit{bound states} for magic state distillation, in analogy with bound states in entanglement distillation~\cite{Horodecki99activate}.  Here we make progress by showing that bound states exist for a very broad class of protocols.   By showing that a single round of a finite sized protocol will not improve these states, it follows that repeating such a protocol, even with an infinite number of iterations, will also have no benefit. Hence, we explain why all known protocols fail to distill some states.

The single-qubit stabilizer states, for which the Gottesman-Knill theorem applies, are the six pure stabilizer states (the eigenstates of $\pm X, \pm Y$ and $\pm Z$) and \earl{any incoherent mixture of these}.   In the Bloch sphere, this convex set with 6 vertices forms the \textit{stabilizer octahedron} partially shown in figure~\ref{fig:Outline}a.  Single-qubit states have density matrices:
\begin{equation}
\label{eqn:NONedge}
 \rho(f, \vec{a}) = \left( \unity + (2f-1)( a_{X} X + a_{Y} Y + a_{Z} Z  ) \right) / 2,
\end{equation}
\earl{where $\vec{a}=(a_{X}, a_{Y}, a_{Z})$ is a unit vector}, and $f$ is the fidelity w.r.t the pure state $|\psi_{\vec{a}}\rangle\langle\psi_\vec{a}|=(\unity + a_{X}X+a_{Y}Y+a_{Z}Z)/2$.  Stabilizer states satisfy:
\begin{equation}
\label{eqn:OctSurface}
	|2f-1|(|a_{X}|+|a_{Y}|+|a_{Z}|) \leq 1
\end{equation}
where the equality holds for states on the surface of the octahedron, and we denote the fidelity of such surface states as ${\fsa}$, which is unique assuming $f \geq 1/2$.

\begin{figure}[t]
\centering
\includegraphics[width=180pt]{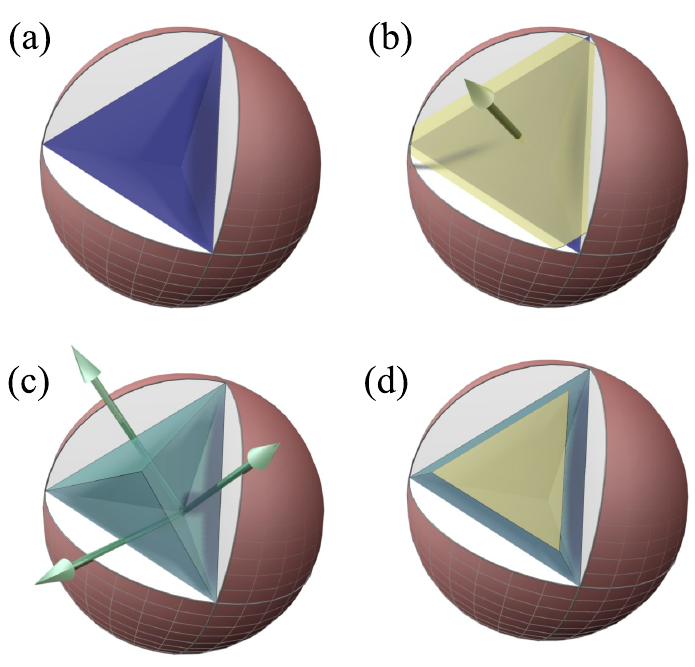}
\caption{One octant of the Bloch sphere with various regions and directions shown. (a) The blue region shows the stabilizer states in one octant.  Each octant is identical, with all stabilizer states forming an octahedron. (b) The yellow plane is the distillation threshold for the 5 qubit code, with the direction of $\vec{a}_{T}$ shown.  The yellow plane is parallel to the underlying blue face of the stabilizer octahedron, but is displaced by a small gap. (c) The three green planes are the thresholds for the Steane code, with each plane differing by local Clifford gates.  These planes meet the stabilizer octahedron at its edges. The three vectors are axes of H-like gates, e.g. $\vec{a}_{H}$; (d) The combined region of states distilled by either the 5 qubit code or the Steane code.  This region only touches the stabilizer octahedron at its edges, and no other known protocol is tight in any other direction.}
\label{fig:Outline}
\end{figure}

Prior protocols for magic state distillation~\cite{BraKit05,Rei01a,Rei02a,Rei03a} increase fidelity towards eigenstates of Clifford gates, such as the Hadamard $H$ and the $T$ gate\footnote{The T gate performs, $TXT^{\dagger}=Y$, $TYT^{\dagger}=Z$.}.  These eigenstates have $\vec{a}_{H}=(1,0,1)/\sqrt{2}$ and $\vec{a}_{T}=(1,1,1)/\sqrt{3}$, with $f=1$ for ideal magic states. Given the ability to prepare a mixed non-stabilizer state, $\rho$, we can perform an operation called \textit{polarization}, or \textit{twirling}, that brings $\rho$ onto a symmetry axis of the octahedron.  For example, by randomly applying $\unity$, $T$ or $T^{\dagger}$, we \earl{map} $\rho \rightarrow \rho(f, \vec{a}_{T})$.

Bravyi and Kitaev proposed the following protocol~\cite{BraKit05} for $\ket{T}$ state distillation: (1) Prepare 5 copies of $\rho(f,\vec{a}_{T})$; (2) Measure the 4 stabilizers of the five-qubit error correcting code; (3) If all measurements give $+1$, the protocol succeeds and the encoded state is decoded into a single qubit state, and otherwise restart.  Upon a successful implementation of this protocol the output qubit has a fidelity $F(f)$ plotted in figure~\ref{fig:fidelities}b.  Provided the initial fidelity is greater than some threshold, a successful implementation yields a higher fidelity.   This protocol has a non-tight threshold, and exhibits a gap between the threshold and the set of stabilizer states.  Because the initial state was twirled onto the T axis, the threshold forms a plane in the Bloch sphere (see figure~\ref{fig:Outline}).  In contrast, Reichardt has proposed a protocol that does have a tight threshold for distillation of $\rho(f, \vec{a}_{H})$ states in a $H$-like direction~\cite{Rei01a}.  His protocol is similar to above, but uses 7 qubits each attempt and measures the 6 stabilizers of the STEANE code~\cite{STEANE95}.  In figure~\ref{fig:fidelities}a we show the performance of this protocol, where there is no threshold gap.  When the initial mixture is not of the form $\rho(f, \vec{a}_{H})$, we twirl the initial mixture onto the $H$ axis.  Hence, the threshold forms a plane for each $H$-like direction (see figure~\ref{fig:Outline}).  Although the protocol is tight in directions crossing an octahedron edge, the protocol fails to distill some mixed states just above the octahedron faces, and so is not tight in all directions.  Even the combined region of states distilled by all known protocols still leaves a set of states above the octahedron faces, whose distillability properties are unknown.  

Here we show that for all size $n$ protocols there is a region of bound states above the octahedron faces.  More formally,  we considering all states $\rho(f, \vec{a}_{P})$ where $\vec{a}_{P}$ has all positive (non-zero) components.  Having all components as non-zero excludes states above octahedron edges. Considering only states in the positive octant is completely general as Clifford gates enable movement between octants.  Many copies of bound states cannot be used to improve on a single copy, and below we formalize the idea of \textit{not improved} and state our main result.

\begin{defin}
We say $\rho'$ is not an improvement on $\rho(f, \vec{a}_{P})$, when $\rho'$ is a convex mixture of $C_{i}\rho(f, \vec{a}_{P})C_{i}^{\dagger}$ and stabilizer states, where $C_{i}$ are Clifford group gates.
\end{defin}

\begin{theorem}
\label{THMgeneralcase}
Consider a device capable of ideal Clifford gates, preparation of stabilizer states, classical feedforward and Pauli measurements.   For any protocol on this device that takes $\rho(f, \vec{a}_{P})^{\otimes n}$ and outputs a single qubit, $\rho'$, there exists an $\epsilon > 0$ such that $\rho'$ is not an improvement on $\rho(f, \vec{a}_{P}) $ for $f \leq {\fsap}+\epsilon$.
\end{theorem}

Theorem~\ref{THMgeneralcase} covers \earl{a wide} class of protocols, \earl{which} attain a fidelity that is upper-bounded by a narrower class of protocols~\cite{Camp09c}, such that theorem~\ref{THMgeneralcase} follows from:

\begin{theorem}
\label{THMstabilizercase}
Consider all protocols that follow these steps: (i) prepare $\rho(f, \vec{a}_{P})^{\otimes n}$; (ii) measure the $n-1$ generators of an $n$ qubit stabilizer code $\mathcal{S}_{n-1}$ with one logical qubit; (iii) postselect on all ``+1" measurement outcomes; (iv)  decode the stabilizer code and output the logical qubit as the single qubit state $\rho'$.    For all such protocols there exists an $\epsilon > 0$ such that $\rho'$ is not an improvement on $\rho(f, \vec{a}_{P}) $ for $f \leq {\fsap}+\epsilon$.
\end{theorem}

\begin{figure}[t]
\centering
\includegraphics[width=\columnwidth]{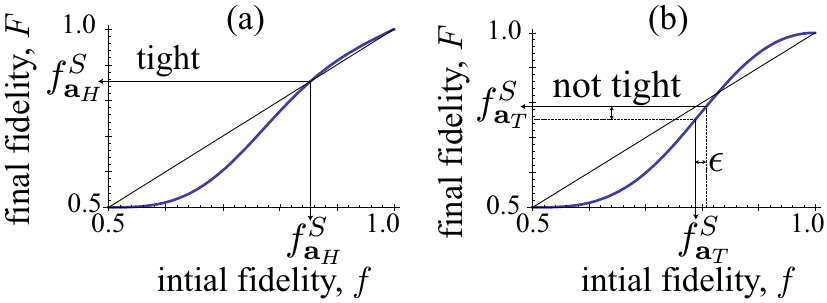}
\caption{The performance of magic state distillation of: (a) the STEANE code for distilling states in a $H$-like direction; (b) the five qubit code distilling states in a $T$-like direction. Notice that both functions are continuous, and that in (b) an input state on the octahedron surface, $f=f^{S}_{\vec{a}_{T}}$, will output a state below the surface.  Consequently, there is also a region above the surface where the output is a stabilizer state, and hence, not an improvement on the initial state.}
\label{fig:fidelities}
\end{figure}

Prior protocols, such as those based on the STEANE code and 5 qubit code, are covered explicitly by theorem~\ref{THMstabilizercase}.  Here we use the structure of stabilizer codes to prove theorem~\ref{THMstabilizercase}, with theorem~\ref{THMgeneralcase} following directly from the results of~\cite{Camp09c}, where such distillation protocols are shown to have equal efficacy with more general Clifford protocols.  It is crucial to consider the implication of these theorems when an $n$-qubit protocol is iterated $m$ times.  When a single round provides no improvement on the initial resource, the input into the second round will only differ by Clifford group operations, and hence our theorem applies to the second, and all subsequent, rounds.  Hence, repeated iteration cannot be used to circumvent our theorem.  Before proving these theorems, we derive a pair of powerful lemmas that identify bound states.  
\begin{lem}
\label{LEM1}
Consider n copies of an octahedron surface state $\rho({\fsap}, \vec{a}_{P})$ projected onto the codespace of $\mathcal{S}_{n-1}$ and then decoded.  If the output qubit is in the octahedron interior, then there exists an $\epsilon >0$ such that for $f\leq {\fsap}+\epsilon$ the same projection on $\rho(f, \vec{a}_{P})^{\otimes n}$ also projects onto a mixed stabilizer state.
\end{lem}
This lemma follows directly from the dependence of the output on $f$, which for finite $n$ is always continuous.  We can observe this lemma at work in figure~\ref{fig:fidelities}b.  Our next lemma identifies when octahedron surface states are projected into the octahedron interior.  Before stating this we must establish some notation.  An initial state,  $\rho({\fsap}, \vec{a}_{P})^{\otimes n}$, is an ensemble of pure stabilizer states:
\begin{equation}
\label{eqn:initial_ensemble}
 \rho({\fsap}, \vec{a}_{P})^{\otimes n}=\sum_{\vec{g} \in \{X, Y, Z\}^{n}} q_{\vec{g}}  \kb{\Psi_{\vec{g}}}{\Psi_{\vec{g}}},
\end{equation}
where $\ket{\Psi_{\vec{g}}}$ is stabilized, $g\ket{\Psi_{\vec{g}}}=\ket{\Psi_{\vec{g}}}$, by the group $\mathcal{G}_{\vec{g}}$ generated by $\vec{g}=$($g_{1}$, $g_{2}$,... $g_{n}$).   The operator $g_{i}$ is $X_{i}$, $Y_{i}$ or $Z_{i}$, with $i$ labeling the qubit on which it acts.  Each contribution has a weighting $q_{\vec{g}} = \prod_{i} (  a_{g_{i}} /(a_{X}+a_{Y}+a_{Z} ) )$.  Measuring the generators of $\mathcal{S}_{n-1}$ and post-selecting on ``+1''  outcomes, projects onto the codespace of $\mathcal{S}_{n-1}$ with projector $P=\sum_{s \in \mathcal{S}_{n-1}}s/2^{n-1} $, producing:
\begin{equation}
\label{eqn:final_state}
 \frac{P  \rho({\fsap}, \vec{a}_{P})^{\otimes n}  P}{\tr [  P  \rho({\fsap}, \vec{a}_{P})^{\otimes n}  P  ]  } =   \sum_{\vec{g} \in \{ X, Y, Z \}^{n} } q'_{\vec{g}}  \kb{\Psi'_{\vec{g}}}{\Psi'_{\vec{g}}},
 \end{equation}
with projected terms, $\ket{\Psi'_{\vec{g}}}$, of new weighting $ q'_{\vec{g}}$.  Each $\ket{\Psi'_{\vec{g}}}$ has its stabilizer generated by ($G_{\vec{g}}, s_{1}, s_{2},.... s_{n-1}$), where $G_{\vec{g}}$ is an independent generator that: (a) was present in the initial group $G_{\vec{g}} \in \mathcal{G}_{\vec{g}}$; and (b) commutes with the measurement stabilizers $G_{\vec{g}} \mathcal{S}_{n-1}= \mathcal{S}_{n-1}G_{\vec{g}}$.  In other words, it must be equivalent to one of six logical Pauli operators of the codespace.  We denote the set of logical operators as  $\mathcal{L}$, and its elements $\pm X_{L}, \pm Y_{L}$ and $\pm Z_{L}$, and so $G_{\vec{g}} \in \mathcal{L} . \mathcal{S}_{n-1}$.  This defines a decoding via the Clifford map, $X_{L}\rightarrow X_{1}$ and $Z_{L} \rightarrow Z_{1}$.  Since there are only six distinct logical states, we can combine many terms in equation~\ref{eqn:final_state}:
\begin{equation} 
 \frac{P  \rho({\fsap}, \vec{a}_{P})^{\otimes n} P}{\tr [  P  \rho({\fsap}, \vec{a}_{P})^{\otimes n} P ] } =   \sum_{L \in \mathcal{L} } q_{L}  \kb{\Psi_{L}}{\Psi_{L}}  ,
\end{equation}
where $\ket{\Psi_{L}}$ has stabilizer generators ($L, s_{1}, s_{2},.... s_{n-1} $).  The new weighting is $q_ {L} =  \sum q'_{\vec{g}}$ with the sum taken over all $\vec{g}$ that generate $\mathcal{G}_{\vec{g}}$ containing an element $G_{\vec{g}}  \in L.\mathcal{S}_{n-1}$.  We can now state the next lemma:

\begin{lem}
\label{LEM2}
Given $n$ copies of $\rho({\fsap}, \vec{a}_{P})$ projected into the codespace of $\mathcal{S}_{n-1}$ and decoded, the output qubit is in the octahedron interior if there exist any two pure states in the initial ensemble, $\ket{\Psi_\vec{g}}$ and $\ket{\Psi_\vec{g'}}$ (defined in equation~\ref{eqn:initial_ensemble}), such that both: \begin{enumerate}
 \item[(i)]  the projected pure states are orthogonal, so that $L \in \mathcal{G}_{\vec{g}}$ and $-s L \in  \mathcal{G}_{\vec{g}'}$ where $L \in \mathcal{L}$ and $s \in \mathcal{S}_{n-1}$; \\
 \textit{and}
 \item[(ii)] upon projection $\ket{\Psi_\vec{g}}$ and $\ket{\Psi_\vec{g'}}$ do not vanish, so $q'_{\vec{g}} \neq 0$ and $q'_{\vec{g}'} \neq 0$.  
\end{enumerate}
\end{lem}
\noindent
We prove this lemma by contradiction.  From equation~\ref{eqn:OctSurface}, and $(2f-1)a_{L}=(q_{L}-q_{-L})$, surface states satisfy:
 \begin{equation}
  |q_{X_{L}}-q_{-X_{L}}|+  |q_{Y_{L}}-q_{-Y_{L}}|+  |q_{Z_{L}}-q_{-Z_{L}}|=1,
\end{equation}
and we assume to the contrary that the projected state has this form.  Since $q_{\pm L}$ are non-negative reals, we have $|q_{L}-q_{-L}|=q_{L}+q_{-L}-2\mathrm{Min}(q_{L}, q_{-L})$, where  $\mathrm{Min}(q_{L}, q_{-L})$ is the minimum of $q_{L}$ and $q_{-L}$.  Along with the normalization condition,  $\sum_{L} q_{L}=1$, this entails:
\begin{equation}
\mathrm{Min}(q_{X_{L}}, q_{-X_{L}})+\mathrm{Min}(q_{Y_{L}}, q_{-Y_{L}})+\mathrm{Min}(q_{Z_{L}}, q_{-Z_{L}}) =0. \nonumber
\end{equation}
Since all terms are positive, no cancellations can occur and so every term must vanish, hence $\mathrm{Min}(q_{L}, q_{-L}) = 0, \forall L$.  However, conditions (\textit{i}) and (\textit{ii}) of the lemma entail that there exists a non-vanishing $\mathrm{Min}(q_{L}, q_{-L})$,  as $q_{L}\geq q'_{\vec{g}} \neq 0$ and $q_{-L}\geq q'_{\vec{g}'} \neq 0$.  Having arrived at this contradiction, we conclude the falsity of the assumption that the projected state remains on the octahedron surface, and so must be in the octahedron interior.  This proves lemma~\ref{LEM2}, and we now show that lemma~\ref{LEM2} applies to all stabilizer reductions that do not trivially take $\rho(f, \vec{a}_{P})^{\otimes n} \rightarrow C_{i} \rho(f, \vec{a}_{P}) C_{i}^{\dagger}$.

Our proof continues by finding canonical generators for the code $\mathcal{S}_{n-1}$.  A related method has been used to prove that all stabilizer states are local Clifford equivalent to a graph state~\cite{VdNDDM01a}, and we review this first.  All stabilizer states have a stabilizer $\mathcal{S}_{n}$ with $n$ generators.   Each generator is a tensor product of $n$ single-qubit Pauli operators.  This can be visualized as an $n$ by $n$ matrix with elements that are Pauli operators, each row a generator and each column a qubit.  Different, yet equivalent, generators are produced by row multiplication, via which we can produce a canonical form.  In this form column $i$ has a non-trivial Pauli operator $A_{i}$ that appears on the diagonal, and all other operators in that column are either the identity or another operator $B_{i}$.  Note that $A_{i}$ and $B_{i}$ compose a third non-trivial Pauli $A_{i}B_{i}=i(-1)^{\gamma_{i}}C_{i}$ with $\gamma_{i}=0,1$.  Hence,  all stabilizer states differ from some graph state by only local Cliffords that map $(A_{i}, B_{i}) \rightarrow (X_{i}, Z_{i})$.

A code, $\mathcal{S}_{n-1}$, has one less generator than the number of qubits, and so more columns than rows.  We can apply the diagonalisation procedure on an $n-1$ by $n-1$ submatrix, to bring this submatrix into canonical form.  Hence, we can find generators of $\mathcal{S}_{n-1}$ such that:
\begin{equation}
  s_{j} = (-1)^{\alpha_{j}} A_{j}  ( \prod_{k \neq j, n} B_{k}^{\beta_{k,j}} ) T_{j, n} ,
\end{equation}
where the variables $\beta_{k,j}=0,1$ denote whether $B_{k}$ or $\unity_{k}$ is present, and $\alpha_{j}=0,1$ defines the phase.  With the $n^{\mathrm{th}}$ column out of canonical form, this leaves the $n^{\mathrm{th}}$ qubit operator $T_{j,n}$ unspecified.  However, if all these generators have $T_{j,n}=\unity_{n}$, then the protocol is trivial and projects $n-1$ qubits into a known stabilizer state and the last qubit untouched, and so no improvement is made for any $f$.  Hence, herein we assume the non-trivial case; in particular we assume stabilizer $T_{n-1, n} \neq \unity_{n}$.  Since, we can always relabel qubits this is completely general.  Furthermore, we can define $T_{n-1, n}=A_{n}$. \earl{Now} we can define a logical operator in the codespace of $\mathcal{S}_{n-1}$:
\begin{eqnarray}
  Z_{L} & = &  \left( \prod_{1 \leq j \leq n-2} B^{\zeta_{j}}_{j} \right)  B_{n-1} B_{n},
\end{eqnarray}
where the variables $\zeta_{j}=0,1$ are uniquely fixed by commutation relations $Z_{L} s_{j}=s_{j} Z_{L}$.  Note that $Z_{L}$ has some inbuilt freedom as $B_{n}$ is not fixed other than that $B_{n}\neq A_{n}, \unity_{n}$, which is equivalent to free choice of $\gamma_{n}$ in the expression $A_{n}B_{n}=i (-1)^{\gamma_{n}}C_{n}$. Now we enquire whether the final state contains two terms stabilized by $Z_{L}$ and $-sZ_{L}$ respectively, hence satisfying the conditions for lemma~\ref{LEM2}.   If we consider the product of $Z_{L}$ and $s_{n-1}$, and choose $\gamma_{n}=\alpha_{n-1} + \gamma_{n-1}$ mod 2, we have:
\begin{equation}
- s_{n-1} Z_{L} =  \left( \prod_{1 \leq k \leq n-2} B_{k}^{\beta_{k, n-1}+\zeta_{k}} \right) C_{n-1}  C_{n} .
\end{equation}
Our choice of $\gamma_{n}$ ensures a minus sign on the left hand side, which aids in finding $\ket{\Psi_{\vec{g}}}$ and $\ket{\Psi_{\vec{g}'}}$  that satisfy our lemma by being stabilized by $G_{\vec{g}} = Z_{L}$ and $G_{\vec{g}'} =-s_{n-1}Z_{L}$ respectively.  This criterion is fulfilled when:
\begin{eqnarray*}
\vec{g} & = & ( B_{1}, B_{2}, .....,  B_{n-2},  B_{n-1},  B_{n}) ,\\ \nonumber
\vec{g'} & = & ( B_{1}, B_{2}, .....,  B_{n-2},  C_{n-1},  C_{n}) .
\end{eqnarray*}
These states only vanish under projection, $q'_{\vec{g}}, q'_{\vec{g'}}=0$, if they are stabilized by the negative of some element of the code $\mathcal{S}_{n-1}$.  To prove they don't vanish, we first observe that every element of $\mathcal{G}_{\vec{g}}$ and $\mathcal{G}_{\vec{g}'}$ has either $\unity_{j}$ or $B_{j}$ acting on qubit $j$, for all $j=1,2,...n-2$.  The only elements of  $\mathcal{S}_{n-1}$ for which this is true are $\unity$ and $s_{n-1}$, but $s_{n-1}$ has $A_{n-1}A_{n}$ acting on the last two qubits and neither $\mathcal{G}_{\vec{g}}$ or $\mathcal{G}_{\vec{g}'}$ contain any such element. 

Using a canonical form of the generators of $\mathcal{S}_{n-1}$, we have shown that non-trivial codes always satisfy the conditions of lemma~\ref{LEM2}.  That is, all non-trivial codespace projections take many surface states into the octahedron interior.  From the continuity expressed by lemma~\ref{LEM1}, this entails the existance of a finite region of non-stabilizer states that are also projected into the octahedron.  Hence, all $n$-copy protocols do no improve on a single copy for some region of bound states above the octahedron faces, completing the proof.  This does not contradict known tight thresholds in edge directions, as these directions have $\vec{a}$ with one zero component.

Although our proof holds for protocols using fixed and finite $n$ copies of $\rho(f, \vec{a}_{P})$, we could conceive of a protocol that varies $n$.  If this varying-$n$ protocol has an $n$-dependent threshold, $f^{T}_{\vec{a}_{P}}(n)$, and $f^{T}_{\vec{a}_{P}}(n) \rightarrow f^{S}_{\vec{a}_{P}}$ as $n \rightarrow \infty$, then its threshold would be arbitrarily suppressible. \earl{Repeated iterations of a protocol, or equivalently employing concatenation of a single-qubit code, will not change the threshold. However, one could consider a broader class of protocols consisting of iterates that act on p qubits and output q qubits (for $p > q > 1$)  followed by a final round outputting a single qubit. Such protocols map $n$ qubits to $1$ qubit, with $n$ growing each iterate, but with only $p$ qubits involved in each iterate.  This implies that multi-qubit output iterates may suppress the threshold effectively, and are worth further study. Currently, no such protocol is known.}  As such, in the asymptotic regime, bound magic states may not exist.  However, numerical evidence so far indicates that smaller codes tend to produce better thresholds than larger codes.  Nevertheless, the theorem does not rule out infinite cases from attaining a tight threshold.  In the regime of finite resources, bound states do exist, and it is interesting ask what  computational power Clifford circuits acting on such states possess.  Can we find methods of efficiently classically simulating bound states; or can bound states be exploited in algorithms that offer a speedup over classical computation? 

Furthermore, our proof assumes a protocol acting on identical copies, which invites study into whether our results extend to non-identical copies.  In particular, following the analogy with entanglement distillation, we speculate that bound magic states may be distillable via ``catalysis'', where some non-consumed distillable resource activates the distillation~\cite{Horodecki99activate}.  Finally we note that noisy Clifford gates can also enable quantum computation~\cite{Plenio08,vanDam09}, and we conjecture that a similar theorem will apply to a class of noisy Clifford gates analogous to states just above the octahedron faces.

The authors would like to thank Shashank Virmani, Matthew Hoban, Tobias Osborne, Ben Reichardt and Steve Flammia  for interesting discussions.  We acknowledge support from the Royal Commission for the Exhibition of 1851, the QIP IRC, QNET and the National Research Foundation and Ministry of Education, Singapore

%\bibliography{references}

\end{document}